\documentclass[12pt,pre,aps,showpacs,preprint]{revtex4}

\usepackage{graphics}
\usepackage{graphicx}
\usepackage{amsfonts}
\usepackage{textcomp}
\usepackage{amssymb}
\usepackage{amsmath}
\usepackage{color}
\usepackage{ulem}

\begin{document}

\title{Robust Algorithm to Generate a Diverse Class of Dense Disordered and Ordered Sphere Packings via Linear Programming}

\author{S. Torquato$^{1,2,3,4,5,6,7}$ and Y. Jiao$^6$}

\affiliation{$^1$Department of Chemistry, Princeton University,
Princeton New Jersey 08544, USA}

\affiliation{$^2$Department of Physics,
Princeton University, Princeton New Jersey 08544, USA}

\affiliation{$^3$Princeton Center for Theoretical Science,
Princeton University, Princeton New Jersey 08544, USA}

\affiliation{$^4$Princeton Institute for the Science and Technology of
Materials, Princeton University, Princeton New Jersey 08544, USA}

\affiliation{$^5$Program in Applied and Computational Mathematics,
Princeton University, Princeton New Jersey 08544, USA}

\affiliation{$^6$Department of Mechanical and Aerospace
Engineering, Princeton University, Princeton New Jersey 08544,
USA}

\begin{abstract}
We have formulated the problem of generating dense packings of
nonoverlapping, non-tiling nonspherical particles  within an adaptive fundamental cell
subject to periodic boundary conditions as an optimization problem
called the Adaptive Shrinking Cell (ASC) formulation [S. Torquato and Y. Jiao,
Phys. Rev. E {\bf 80}, 041104 (2009)]. Because the objective function and impenetrability
constraints can be exactly linearized for sphere packings with a size distribution
in $d$-dimensional Euclidean space $\mathbb{R}^d$, it is most suitable and natural
to solve the corresponding ASC optimization problem using
sequential linear programming (SLP) techniques. We implement an SLP solution
 to produce robustly a wide spectrum
of jammed sphere packings in $\mathbb{R}^d$ for $d=2,3,4,5$ and $6$
with a diversity of disorder and densities up to the respective maximal densities.
A novel feature of this deterministic algorithm is that it can produce
a broad range of inherent structures (locally maximally dense
and mechanically stable packings), besides the usual
disordered ones (such as the maximally random jammed state), with very small computational cost
compared to that of the best known packing algorithms
by tuning the radius of the {\it influence sphere}.
For example, in three dimensions, we show that it can produce
with high probability a variety of
strictly jammed packings with a packing density anywhere in the wide range $[0.6, 0.7408\ldots]$,
where $\pi/\sqrt{18}=0.7408 \ldots$ corresponds to the density of
the densest packing. We also apply the algorithm to generate various disordered packings
as well as the maximally dense packings for $d=2,4,5$ and 6.
Our jammed sphere packings are characterized
and compared to the corresponding packings generated by the well-known
Lubachevsky-Stillinger molecular-dynamics packing algorithm.
Compared to the LS procedure, our SLP protocol is able to
ensure that the final packings are truly jammed,
produces disordered jammed packings with anomalously low
densities, and is appreciably more robust and computationally
faster at generating maximally dense packings, especially as the
space dimension increases.

\end{abstract}

\maketitle

\section{Introduction}

Hard-particle packings have provided a rich source of outstanding theoretical problems and served
as useful starting points to model the structure of  liquids \cite{Ha86,Chaik00},
 glasses \cite{Za83,To00,To02a,Pa10}, crystals \cite{Sa62,To07,As08}, colloids \cite{Ru89,Chaik00},
granular media \cite{Me94,
Ed01,To01,To10}, living cells \cite{To02a,Ge08}, random media \cite{To02a,Zo06,Br07, ToMater}, 
and polymeric systems \cite{Kara08, Fote08}.
We focus our attention in this paper on the venerable
idealized hard-sphere model in $d$-dimensional Euclidean space $\mathbb{R}^d$ in
which the only interparticle interaction
is an infinite repulsion for overlapping particles, which can be
thought of as the ``Ising model" for hard spheres \cite{To10}.

There has been resurgent
interest in hard-sphere packings in dimensions
greater than three in both the physical and mathematical sciences.
For example, it is known that the optimal way of
sending digital signals over noisy channels corresponds to
the densest sphere packing in a high-dimensional space \cite{Sh48,Co93}.
These error-correcting codes underlie a variety of systems
in digital communications and storage, including compact
disks, cell phones and the Internet. Physicists have studied
hard-sphere packings in high dimensions to gain insight into
ground and glassy states of matter as well as phase behavior
in lower dimensions \cite{Fr99,Pa00,To06,Sk06,To06b,Ch09}. The determination of the densest
packings in arbitrary dimension is a problem of longstanding
interest in discrete geometry and number theory \cite{Co93,Co03,Co09}.

The hard-sphere Ising model enables us to be precise about the important concept of
``jamming." The generation and characterization of jammed packings are the
main concerns of this paper.
A jammed packing is one in which the particle positions
are fixed by the impenetrability constraints and boundary
conditions (e.g., hard-wall or periodic boundary conditions).
For sphere packings, one can define  hierarchical jamming categories
beginning from the least restrictive to the most stringent
one \cite{To01}. In particular, a packing is {\it locally} jammed if
no particle in the system can be translated while fixing the
positions of all other particles, requiring that each sphere in the packing
be in contact with at least $d+1$ spheres not all in the same
hemisphere. A {\it collectively} jammed packing is a locally jammed
packing such that no subset
of spheres can simultaneously be continuously displaced
so that its members move out of contact with one another
and with the remainder set. A packing is {\it strictly} jammed if it is collectively
jammed and all globally uniform volume nonincreasing
deformations of the system boundary are disallowed
by the impenetrability constraints. The reader is referred to
Refs. \cite{To01} and \cite{To10} for further details.
Our main concern in this paper will be packings
that are at least collectively jammed.

The packing density $\phi$, the fraction of $\mathbb{R}^d$
covered by the spheres, is the simplest characteristic
of a jammed packing. However, such a characterization is clearly not
sufficient in order to distinguish between ordered and disordered
packings. In fact, jammed packings may be produced with
variable degrees of disorder/order. It has been suggested
that a scalar order metric $\psi$ be employed to measure the
degree  of order in a packing,
such that $\psi =1$ corresponds to fully ordered [e.g.,  the
perfect face-centered cubic (fcc) crystal in three dimensions]
 and $\psi =0$ corresponds to perfectly a disordered
Poisson distribution of sphere centers \cite{To00,Ka02}.
Thus, very large jammed packings have been classified based on their locations
in the density-order $\phi$-$\psi$ plane, called the {\it order map}.
The points on the boundary of the jammed domain in the order map
constitute optimal packings. For example, in $\mathbb{R}^3$,
the maximally dense fcc lattice sphere packing ($\phi=\pi/\sqrt{18} =0.7048 \ldots$)
\cite{Ha05} also has the highest order metric ($\psi=1$) when appropriately defined.
Another extremal state of special interest is the
maximally random jammed (MRJ) state,  which is the one that minimizes
a scalar order metric $\psi$ subject to the condition of the
degree of jamming \cite{To00,To10}. Studies of different
order metrics \cite{Tr00} for three-dimensional frictionless spheres have consistently led to
a minimum  at approximately the same density $\phi \approx 0.64$
for collective and strict jamming in the order map \cite{To00,Ka02}.
Such MRJ packings have also been shown to be hyperuniform, i.e.,
infinite-wave-length density fluctuations vanish \cite{Do05c}.
The lowest density strictly jammed states in $\mathbb{R}^3$, thought to be tunneled crystals \cite{To07},
are a fascinating set of extremal loci  in the jamming region of the order map \cite{To10}.
The frequency of occurrence
of a particular configuration is irrelevant insofar as the order
map is concerned, i.e., the order map emphasizes a
``geometric-structure'' approach to analyze packings by
characterizing individual configurations, regardless of their
occurrence probability \cite{To10}.

During the last two decades, the Lubachevsky-Stillinger (LS) algorithm \cite{Lu90} has been
the premier workhorse to generate
a wide spectrum of dense jammed sphere packings
with variable disorder in both two and three dimensions \cite{To00,Ka02,Tr00}.
This is an event-driven (or collision-driven) molecular dynamics algorithm
in which an initial configuration of spheres of a given size within
a periodic box are given initial random velocities and
the motion  of the spheres are followed as they collide elastically and also
expand uniformly until the spheres can no longer expand.
This algorithm has been generalized by Donev, Torquato and Stillinger \cite{Do05a}
to generate jammed packings of smoothly-shaped nonspherical particles,
including ellipsoids \cite{Do07}, superdisks \cite{Ji08} and superballs \cite{Ji09}.

Not surprisingly, this packing protocol is not without some
inadequacies. Event-driven packing protocols with growing
particles do not guarantee jamming of the final packing
configuration, since jamming is not explicitly incorporated as a 
termination criterion. To produce jammed random packings, for
example, a large expansion rate is necessary in the early stages
of the simulation to suppress crystallization, but using a  high
expansion rate is highly undesirable toward the end of the
simulation, which often  leads to unjammed configurations. Thus
either a variable expansion rate must be used (which decreases as
a function time in some arbitrary fashion) or, if a uniform
expansion is employed, the spheres of the terminal packing must be
shrunk by some arbitrary small amount  and then this initial
packing configuration must be re-densified using a very small
expansion rate. Maximally dense jammed packings are highly
computationally expensive to generate, especially in high
dimensions,  because a very small expansion rate is necessary (on
the order of $10^{-6}$ to $10^{-10}$ \cite{rate}, dependent on
dimension). We will see that in five and six dimensions, even a
expansion rate of $10^{-10}$ fails to produce the maximal-density
packings. Moreover, a large number of total collisions  per
particle is required; for MRJ packings and the densest known
maximally-dense packings, on the order of $10^5$ and $10^{7}$
collisions per particle are required, respectively, the latter of
which is computationally very costly. Finally, such event-driven
packing protocols evolve stochastic velocity initializations,
which makes one have less control of the final packings via the
initial configuration.

The next generation packing protocol to generate jammed sphere
packings should retain the versatility and advantages of the
LS algorithm while correcting its imperfections, including
improving computational speed.
We show that our proposed sequential linear programming (SLP) solution
of the Adaptive Shrinking Cell (ASC) optimization problem
formulated elsewhere for general particle shapes
(including polyhedra) \cite{To09}  indeed has all of these desirable features in so far
as jammed sphere packings with a size distribution are
concerned. Because the design variables (including
periodic simulation box shape and size) and impenetrability constraints
can be exactly linearized for spheres packings,
the deterministic SLP solution in principle always leads to jammed packings
(up to a high numerical tolerance) with a wider
range of densities and degree of disorder than packings produced by the LS algorithm.
Each linear programming solution step starting from some
initial particle configuration involves a deterministic collective
motion of the entire particle configuration to a
higher density and, because the periodic simulation box
can simultaneously deform and shrink, the final
state is guaranteed to be at least collectively jammed in principle.
Whereas the LS algorithm requires between $10^5$ and $10^{7}$
collisions per particle without ensuring jammed final states,
the deterministic SLP solution, which is easy to implement, requires only 10 to 100
linear-programming steps to achieve jammed packings,
depending on the desired density, which can be controlled
by tuning what we call the size of the {\it influence
sphere}. Thus, the SLP algorithm is computationally very efficient
in generating the maximally dense packings, even in
high space dimensions. By appropriately choosing
initial conditions, one can achieve strictly jammed
disordered sphere packings with anomalously low densities, e.g.
$\phi \approx 0.6$ in three dimensions \cite{pnas,pnas-f}.
Indeed, a novel feature of the algorithm is that it can produce
a broad range of {\it inherent structures} (locally maximally dense
and mechanically stable packings), such as the maximally random jammed states
as well as the globally maximally dense inherent structures, with very small computational cost.

 The rest of the paper is organized as follows. In Sec.~II, we provide
basic definitions for packing problems. In Sec.~III, we present the mathematical
formulation and algorithmic details of our sequential-linear-programming (SLP)
procedure to solve the adaptive-shrinking-cell (ASC) optimization problem
to generate jammed sphere packings.
In Sec.~IV, we discuss the energy landscape (negative of the packing density)
for jammed sphere packings (inherent structures)
and show that our SLP solution procedure is able to lead to both the
mechanically stable local minima  (i.e., local density maxima) and the global minima (maximal-density packings). In Sec.~V, we employ
the SLP algorithm to produce a diverse class of disordered
jammed packings, including MRJ states, as well as
maximal-density packings of hard spheres in $\mathbb{R}^d$ for $d=2, 3, 4, 5$ and $6$. The
characteristics of these packings are compared to those generated using the LS algorithm.
In Sec.~VI, we employ the SLP algorithm to produce jammed packings in $\mathbb{R}^3$
with varying degrees of disorder
and packing densities anywhere in the wide range $[0.6, 0.7408\ldots]$.
In Sec.~VII, we discuss the ramification of our results and make concluding remarks.


\section{Definitions}
\label{defs}

A {\it lattice} $\Lambda$  in $\mathbb{R}^d$ is a subgroup
consisting of integer linear combinations of vectors $\boldsymbol{\lambda}_i$ that constitute
a basis for $\mathbb{R}^d$.
In a lattice, the space $\mathbb{R}^d$ can be geometrically divided into identical
regions $F$ called {\it fundamental cells}, each of which contains
one lattice site specified by the {\it lattice vector}
\begin{equation}
{\bf p}= n_1 {\boldsymbol \lambda}_1 + n_2 {\boldsymbol \lambda}_2+ \cdots + n_{d-1} {\boldsymbol \lambda}_{d-1}+n_d {\boldsymbol \lambda}_d,
\end{equation}
where $\boldsymbol{\lambda}_i$ ($i=1,2,\ldots,d$) are the {\it basis vectors}
and  $n_i$ spans all the integers for $i=1,2,\cdots d$.
The {\it generator matrix} ${\bf M}_\Lambda=\{\boldsymbol{\lambda}_1,\boldsymbol{\lambda}_2,\ldots, \boldsymbol{\lambda}_d\}$ of a lattice $\Lambda$ is a matrix
with the basis vectors $\boldsymbol{\lambda}_i$ as columns and, thus, contains $d^2$ elements.
In the physical sciences and engineering, a lattice $\Lambda$ is referred to as a {\it Bravais}
lattice. Unless otherwise stated, the term ``lattice" will refer here
to a Bravais lattice only.
A {\it lattice packing} of congruent spheres $P_L$ is one in which  the centers of the nonoverlapping identical
spheres are located at the points of $\Lambda$
and hence each fundamental cell contains within it exactly one sphere.
Thus, the density of a lattice packing of spheres
of radius $R$ is given by
\begin{equation}
\phi= \frac{v_1(R)}{v_F},
\end{equation}
where $v_F$ is the volume of the fundamental cell and
\begin{equation}
v_1(R)=\frac{\pi^{d/2} R^d}{\Gamma(1+d/2)},
\label{vol-sph}
\end{equation}
is the volume of  a $d$-dimensional spherical particle of radius $R$
and $\Gamma(x)$ is the Euler gamma function.

\begin{figure}[htp]
\begin{center}
$\begin{array}{c@{\hspace{0.05cm}}c}
\includegraphics[height=6cm, keepaspectratio]{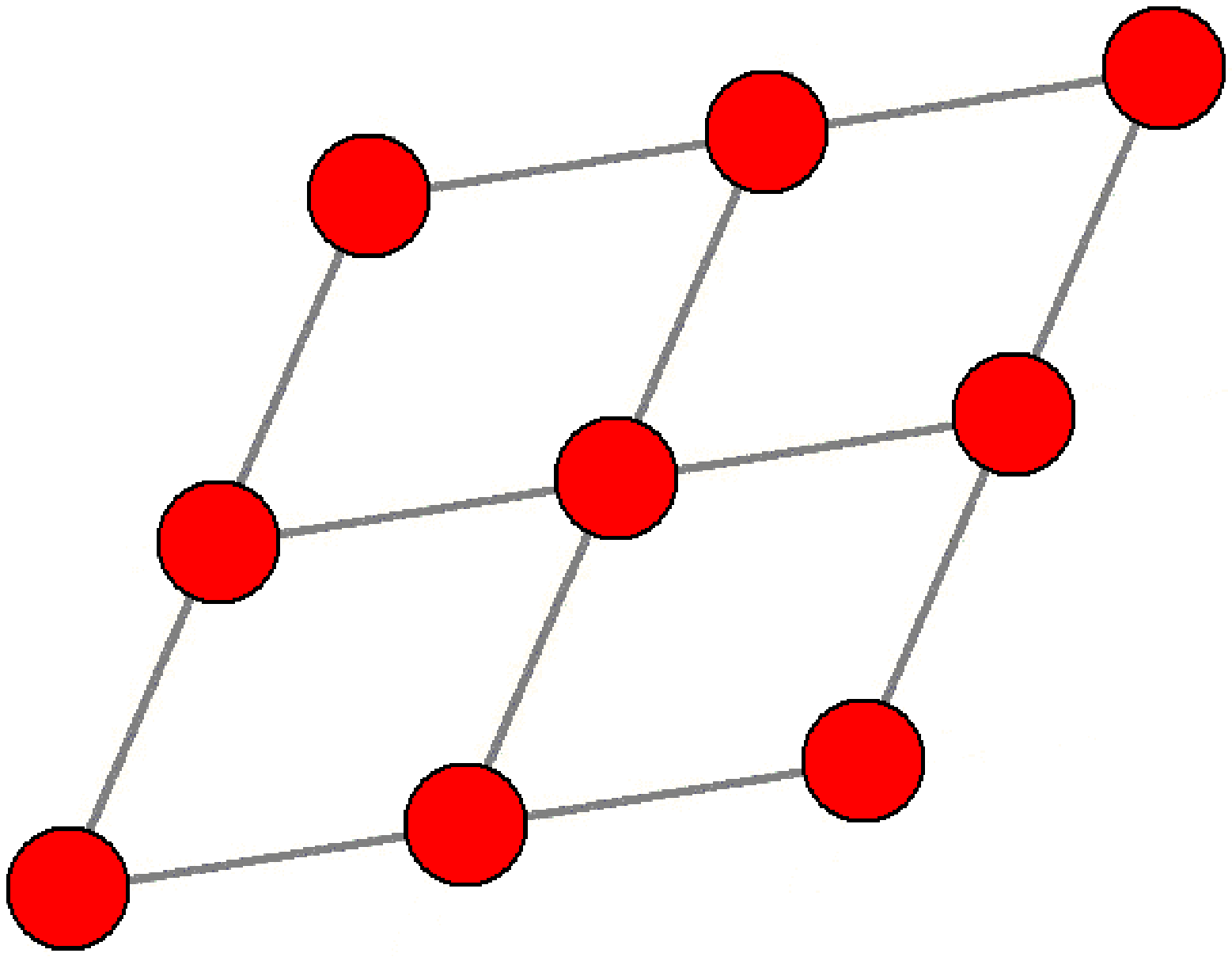} &
\includegraphics[height=6cm, keepaspectratio]{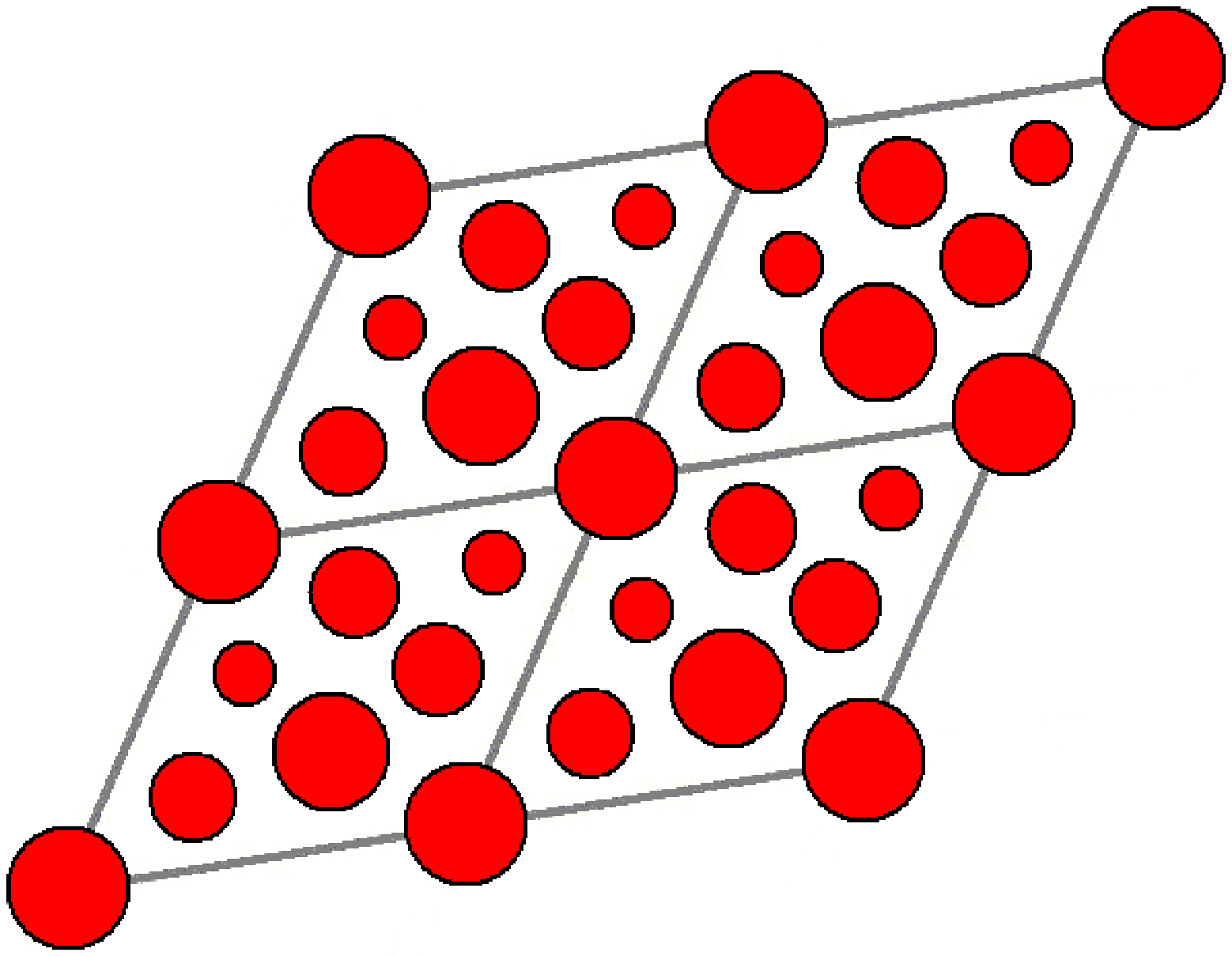} \\
\mbox{(a)} & \mbox{(b)} \\\\
\end{array}$
\end{center}
\caption{(Color online). Illustrations of two basic
sphere packings in $\mathbb{R}^2$: (a) lattice packing of
circular hard disks in $\mathbb{R}^2$, which necessarily can only involve congruent
circles, and (b) periodic packing of circles of different sizes. }
\label{fig1}
\end{figure}

A more general notion than a lattice packing is a periodic
packing of spheres (not necessarily identical in size), which is obtained by placing
a fixed configuration of $N$ spheres (where $N\ge 1$) of radii $R_1, R_2, \ldots, R_N$ in one
fundamental cell of a lattice $\Lambda$, which is then periodically replicated
without overlaps.
 Thus, the packing is still
periodic under translations by $\Lambda$, but the $N$ spheres can occur
anywhere in the chosen fundamental cell subject to the overall nonoverlap condition.
The packing density of a  periodic packing of spheres is given by
\begin{equation}
\phi=\frac{\sum_{i=1}^N v_1(R_i)}{v_F}=\rho \langle v_1(R) \rangle,
\end{equation}
where $\rho=N/v_F$ is the {\it number density}, i.e., the number of particles per unit volume, and
\begin{equation}
 \langle v_1(R) \rangle= \frac{1}{N} \sum_{i=1}^N v_1(R_i)
\label{den-per}
\end{equation}
expected volume of a sphere.
Figure \ref{fig1} illustrates in $\mathbb{R}^2$  a lattice packing of spheres (which by definition
requires that they be identical is size) and a periodic packing of spheres in which
the spheres have a size distribution.

\section{Mathematical and Algorithmic Details}

We have formulated the problem of generating dense packings of
nonoverlapping, non-spherical particles within a fundamental cell in
$\mathbb{R}^3$
subject to periodic boundary conditions as an optimization problem
called the the Adaptive Shrinking Cell (ASC) scheme \cite{To09}.
The objective function
is taken to be the negative of the packing density $\phi$. Starting from
an initial unsaturated packing
configuration of particles of fixed size in the fundamental cell, the
positions and orientations
of the particles are design variables for the optimization.
We also allow the boundary of the fundamental cell
to deform macroscopically as well as compress or expand (while
keeping the particles fixed in size) such that there is a net compression
(increase of the density of the packing) in the final state. Therefore,
the deformation and compression/expansion of the cell
boundary, called the {\it adaptive fundamental cell},
are also design variables.

The ASC optimization problem can be solved using
various techniques, depending on the shapes of the particles. For example,
for polyhedron particles the nonoverlap condition is highly nonlinear,
which makes it inefficient to
solve even using nonlinear programming methods, and hence Monte Carlo
techniques provide an efficient means of solving the problem [43]. However, for spheres,
one can exactly linearize the
objective function, design variables and constraints, enabling one to
exploit efficient linear programming techniques.

We first explicitly formulate the ASC scheme for noncongruent sphere
packings, i.e., spheres with a {\it size distribution} in $\mathbb{R}^d$.
(Obviously, for sphere packings, particle rotations are irrelevant.)
Then we describe the solution of the ASC optimization problem using
sequential linear programming (SLP) techniques.

\subsection{Adaptive Shrinking Cell (ASC) Scheme for Spheres}

Consider a periodic packing of $N$ spheres with diameters $D_1,D_2,\ldots,
D_N$ in a fundamental cell $F$ of a lattice $\Lambda$ in $\mathbb{R}^d$.
As noted in Sec.~\ref{defs}, the fundamental cell is specified
by the generator matrix ${\bf
M}_\Lambda=\{\boldsymbol{\lambda}_1,\boldsymbol{\lambda}_2,\ldots,
\boldsymbol{\lambda}_d\}$, where $\boldsymbol{\lambda}_i$ are the
lattice basis vectors. The design variables of the ASC scheme includes
both particle displacements, and deformation and volume change of the fundamental cell $F$.
Let the {\it global} position of the center of sphere $n$ (referred to
some arbitrary set of non-orthogonal basis vectors) be denoted by ${\bf x}^g_n$.
For simplicity, it is useful to choose orthogonal bases vectors or a
Cartesian system. Denote by $\Delta {\bf x}^g_n$ a translational displacement of this
particle center. Thus, the new position of the center of particle $n$
is given by
\begin{equation}
\tilde{\bf x}^g_n = {\bf x}^g_n+\Delta {\bf x}^g_n.
\label{eq1}
\end{equation}
However, since the fundamental cell is adaptive, we also need
to consider the positions of the particle centers in terms
of the lattice basis vectors $\boldsymbol{\lambda}_i$ ($i=1,2,\ldots,d$).
For the center of particle $n$, this
relative position ${\bf x}^\lambda_n$ is related to the global position
${\bf x}^g_n$ by the transformation \cite{footnote2}
\begin{equation}
{\bf x}^\lambda_n = {{\bf M}_\Lambda}^{-1}\cdot{\bf x}^g_n.
\label{eq3}
\end{equation}

The adaptive fundamental cell allows for a small change in the fundamental
cell
\begin{equation}
\Delta {\bf M}_\Lambda = {\boldsymbol\varepsilon}\cdot{\bf M}_\Lambda,
\label{eq4}
\end{equation}
including both volume and shape changes, where ${\boldsymbol \varepsilon}$
is a symmetric strain tensor, i.e.,
\begin{equation}
{\boldsymbol\varepsilon} = \left [{
\begin{array} {c@{\hspace{0.35cm}}c@{\hspace{0.35cm}}c@{\hspace{0.35cm}}c}
\epsilon_{11} & \epsilon_{12} & \ldots & \epsilon_{1d} \\
\epsilon_{21} & \epsilon_{22} & \ldots & \epsilon_{2d} \\
\vdots & \vdots & \vdots & \vdots \\
\epsilon_{d1} & \epsilon_{d2} & \ldots & \epsilon_{dd} \end{array}
}\right ],
\end{equation}
The new lattice is specified by the new matrix generator
\begin{equation}
\tilde{\bf M}_\Lambda = {\bf M}_\Lambda + \Delta {\bf M}_\Lambda.
\label{eq5}
\end{equation}
Substituting the above equation into Eq.~(\ref{eq3}) yields
\begin{equation}
{\bf \tilde{x}}^g_n = \tilde{\bf M}_\Lambda \cdot{\bf x}^\lambda_n = {\bf
x}^g_n+\Delta {\bf M}_ \Lambda\cdot{\bf x}^\lambda_n.
\label{eq6}
\end{equation}

Straining the fundamental cell corresponds to non-trivial
\textit{collective} motions of the particle centers.
In general, the translational motions of the particles contain
contributions from a direct part [given by Eq.~(\ref{eq1})]
and the collective motion imposed by the adaptive fundamental cell. It is
this
collective motion that enables the algorithm to explore the
configuration space more efficiently and to produce highly dense packings.

The displacement ${\bf r}^\lambda_{mn}$ pointing from the centroid of
sphere $m$ to that of sphere $n$ in terms lattice basis vectors
$\boldsymbol{\lambda}_i$ is given by
\begin{equation}
{\bf r}^\lambda_{mn} = {\bf x}^\lambda_n - {\bf x}^\lambda_m,
\end{equation}
and the counterpart global displacement vector ${\bf r}^g_{mn}$ is
\begin{equation}
{\bf r}^g_{mn} = {\bf x}^g_n - {\bf x}^g_m.
\end{equation}
The Euclidean distance ${r}^g_{mn}$ between the centroids is then given by
\begin{equation}
r^g_{mn} = |{\bf r}^g_{mn}| = \sqrt{{\bf r}^g_{mn}\cdot {\bf r}^g_{mn}} =
\sqrt{{\bf
r}^\lambda_{mn}\cdot {\bf G} \cdot {\bf r}^\lambda_{mn}},
\end{equation}
where ${\bf G} = {\bf M}_\Lambda^T \cdot {\bf M}_\Lambda$ is the
\textit{Gram matrix} of the lattice $\Lambda$. Thus, the general
mathematical
formulation of the our ASC optimization scheme for hard sphere
packings in $d$-dimensional Euclidean space $\mathbb{R}^d$ is:
\begin{equation}
\begin{array}{c}
\!\!\!\!\!\!\!\!\!\!\!\!\!\!\!\!\!\!\!\!\!\!\!\!\!\!\!\!\!\!\!\!\!\!\!\!\!\!\!\!
\!\!\!\!\!\!\!\!\!\!\!\!\!\!\!\!\!\!\!\!\!\!\!\!\!\!\!\!\!\!\!\!\!\!\!\!\!\!\!\!\!\!\!\!
\mbox{minimize}~ -\phi({\bf x}^{\lambda}_1,\ldots,{\bf x}^{\lambda}_N; {\bf M}_\Lambda) \\
\mbox{subject to:}~ r^g_{mn} \ge \overline{D}_{mn}, ~\mbox{for all
neighbor
pairs}~ (m,n)~ \mbox{of interest}.
\end{array}
\end{equation}
where $\overline{D}_{mn} = (D_m+D_n)/2$.

It should be emphasized
that the {\it neighbor} pairs here do not necessarily mean nearest
neighbors. Instead, they are determined by a  distance
$\gamma_{mn}$, i.e., two spheres $m$ and $n$ are neighbors of one
another if their pair distance $r^g_{mn}<\gamma_{mn}$. Thus, by
{\it near neighbors} of a given sphere we mean all of the spheres
within some radius of that given sphere. Here, we choose
$\gamma_{mn}=\alpha\overline{D}_{mn}$ and $\alpha$ is a positive
in the interval $[1,L/2\overline{D}_{mn}]$,
where $L$ is the length of the shortest lattice vector associated
with the fundamental cell. For
monodisperse packings, $\gamma_{mn}$ is identical for all particle
pairs $(m, n)$. For polydisperse packings, $\gamma_{mn}$ is generally
different for each pair $(m, n)$. We call $\gamma_{mn}$ the radius of
{\it influence sphere} associated with each pair $(m, n)$.
As we will see, these radii are crucial in determining
the density and degree of disorder of the final jammed states.

\subsection{Solving the ASC formulation using Sequential Linear
Programming}

The formulated problem for the ASC scheme can be solved by
considering an equivalent \textit{sequential linear programming}
(SLP) problem. Suppose that we only allow the fundamental cell to
change by a small amount from its original size and shape, and only allow
small particle displacements; then both the impenetrability constraints and the
objective function can be linearized. Due to the linearization, one needs now to
explicitly places bounds on the strain components and the particle displacements,
which do not appear in the original problem. Once the linearized problem is
solved via the linear programming method, a new many-particle configuration
can be generated using the resulting particle
displacements and  fundamental cell. Then this
new configuration is used as an initial configuration, based on which a
new LP problem is formulated and solved. This sequential
linear programming procedure is repeated until the increase
of the packing density is smaller than some prescribed small tolerance
value, implying the system is jammed up to high
numerical accuracy.


Consider an initial packing configuration of $N$ spheres within a fundamental
cell of volume $v^0_F$. According to relation (4), the density of the packing is given by
\begin{equation}
\phi_0 = \frac{\sum_{i=1}^N v_1(R_i)}{v^0_F},
\end{equation}
where $v_1(R_i)$ is the volume of sphere $i$ with radius $R_i$. Now consider
a small change of the fundamental cell that leads to a cell
volume change $\Delta v$, as well as a change of the
positions of the centers of every sphere that obey
the impenetrability constraint. The density of the new configuration
in the new fundamental cell is given by
\begin{equation}
\phi = \frac{\sum_{i=1}^N v_1(R_i)}{v^0_F+\Delta v}.
\end{equation}
If $\Delta v$ is sufficiently small, we can expand $\phi$ as a Taylor series
in $\Delta v$, keeping only the linear terms, i.e.,
\begin{eqnarray}
\phi &=& \frac{\sum_{i=1}^N v_1(R_i)}{v^0_F+\Delta v} \nonumber\\
&\approx& \frac{\sum_{i=1}^N v_1(R_i)}{v^0_F}\left [{1-\frac{\Delta v}{v^0_F}}\right] \nonumber \\
&=& \phi_0 [1- tr(\boldsymbol\varepsilon)],
\end{eqnarray}
where we have used the relation $\Delta v/v^0_F = tr(\boldsymbol\varepsilon)$,
i.e., the relative small volume change is given by the trace of the strain tensor.
We see  from  the above equation that maximizing the packing density $\phi$ is
equivalent to minimizing the {\it trace of the strain tensor} $tr(\boldsymbol\varepsilon)$.

The nonoverlapping conditions can be linearized in a similar way, which leads
to the following sequential linear programming formulation for the ASC procedure:
\begin{equation}
\begin{array}{c}
\!\!\!\!\!\!\!\!\!\!\!\!\!\!\!\!\!\!\!\!\!\!\!\!\!\!\!\!\!\!\!\!\!\!\!\!\!\!\!\!
\!\!\!\!\!\!\!\!\!\!\!\!\!\!\!\!\!\!\!\!\!\!\!\!\!\!\!\!\!\!\!\!
\mbox{minimize}~ tr(\boldsymbol\varepsilon) =
\epsilon_{11}+\epsilon_{22}+\cdots+\epsilon_{dd} \\
\!\!\!\!\!\!\!\!\!\!\!\!\!\!\!\!\!\!\!\!\!\!\!\!\!\!\!\!\!\!\!\!\!\!\!\!\!\!\!\!
\!\!\!\!\!\!\!\!\!\!\!\!\!\!\!\!\!\!\!\!\!\!\!\!\!\!\!\!\!\!\!\!\!\!\!\!\!\!\!\!\!\!\!\!\!\!\!\!
\!\!\!\!\!\!\!\!\!\!\!\!\!\!\!\!\!\!\!\!\!\!\!\!\!\!\!\!\!\!\!\!\!\!\!\!\!\!\!\!\!\!\!\!\!\!\!\!
\!\!
\mbox{subject to:}~ \\
{\bf M}_\Lambda \cdot {\bf r}^\lambda_{nm} \cdot\epsilon\cdot {\bf
M}_\Lambda \cdot
{\bf r}^\lambda_{nm} + \Delta{\bf x}^\lambda_m\cdot {\bf G} \cdot {\bf
r}^\lambda_{nm}+ \Delta{\bf x}^\lambda_n\cdot {\bf G} \cdot {\bf
r}^\lambda_{mn} \\
\ge \frac{1}{2}(\overline{D}_{mn}^2 - {\bf r}^\lambda_{nm}\cdot {\bf
G}\cdot {\bf r}^\lambda_{nm})+{\cal R}, \\
~\mbox{for all neighbor pairs}~ (m,n)~ \mbox{of interest,} \\
\!\!\!\!\!\!\!\!\!
\Delta{\bf x}^{\lambda, lower}_n \le \Delta{\bf x}^\lambda_n \le \Delta{\bf
x}^{\lambda, upper}_n, ~\mbox{for all} ~n = (1, ..., N), \\
\!\!\!\!\!\!\!\!\!\!\!\!\!\!\!\!\!\!\!\!\!\!\!\!\!\!\!\!\!\!\!\!\!\!\!\!\!\!\!\!
\!\!\!\!\!\!\!\!\!\!\!\!\!\!\!\!\!\!\!\!\!\!\!\!\!\!\!\!\!\!\!\!\!\!\!\!\!\!\!\!\!\!\!
\!\!\!\!\!\!\!\!\!\!
{\boldsymbol\varepsilon}^{lower}\le {\boldsymbol\varepsilon}\le
{\boldsymbol\varepsilon}^{upper}.
\end{array}
\label{LP}
\end{equation}
Note that the tensor/vector inequalities in (\ref{LP}) apply to the
corresponding components
and ${\bf r}^\lambda_{nm} = -{\bf r}^\lambda_{mn} =
{\bf x}^\lambda_m - {\bf x}^\lambda_n$ is the displacement vector
between spheres $m$ and $n$ in the initial configuration.
The scalar quantity ${\cal R}>0$ is a relaxation variable, which
accounts for the effects of higher-order terms that would be ignored
in the purely linearized problem. The value of ${\cal R}$ is determined by
the bounds on the design variables, i.e.,
\begin{equation}
{\cal R} = max_{(m,n)}\{-\Delta {\bf M}_\Lambda\cdot \Delta{\bf
x}^\lambda_{mn}\}.
\end{equation}
Introducing ${\cal R}$ has several advantages. Firstly, it makes the
linearized problem rigorously
equivalent to the original (unlinearized) optimization problem. Secondly,
it enables us to set practically large bounds on the design
variables (e.g., the strain components and the particle displacements),
which in turn enables the algorithm to explore a larger region
of the configuration space around the initial point and makes it more
efficient to generate dense packings. In practice, however, we find that as
long as the bounds on the design variables are chosen
to be sufficiently small, one can safely set ${\cal R} = 0$ and check the
generated packing to ensure that no impenetrability constraints are
violated. If some of them are violated due to the deformable 
fundamental cell, i.e., leading to a non-positive-definite quadratic 
term associated with the strain tensor in these conditions, the bound widths are
reduced to one half of their original value and the linearized problem is
re-solved.

Since the objective function and constraints in
the new problem are linear functions of the design variables, the new
problem (\ref{LP}) is solved using standard linear programming methods
(e.g., simplex method, interior-point method, etc.).
Starting from an initial packing configuration, we solve the
linearized problem and find a new packing, which is denser and
slightly different from the starting configuration. Then the new
packing is used as the starting configuration, and a new
linearized problem is solved. By repeating this process, one
actually solves a sequence of linear programming problems to generate a denser packing
from the previous configuration.

The choice of the bounds for the design variables is also
important in practice. For example, different bound widths for the
principle strain components and shear strain components correspond
to different compression and deformation rates for the packing.
Choosing the bound widths carefully can dramatically improve the
ability of the algorithm to generate dense packings. We also emphasize
that the choice of neighbor pairs (i.e., the value of the
influence sphere radius $\gamma_{mn}$) is also nontrivial, which we
will elucidate in the ensuing discussion.

\subsection{Ensuring Jamming}

Finally, we note that our SLP
solution procedure guarantees (up to numerical precision)
that the final locally or globally maximally dense states
are indeed strictly jammed under periodic boundary conditions.
In general, each impenetrability (nonoverlapping) condition
defines a curved hypersurface in the configuration space that
separates accessible and inaccessible regions \cite{Do05_g2, To10}. Linearization
of the impenetrability conditions corresponds to replacing the
curved hypersurfaces with hyperplanes, which further reduces
the accessible region (i.e., the linearized conditions are even
stronger than the original conditions). However, near the jamming
point, the available configuration space for the spheres
asymptotically approaches a closed convex polytope which
is exactly determined from the linearized impenetrability conditions \cite{Do05_g2, To10}.
In other words, toward the jamming point, the linearized
conditions become exact nonoverlapping conditions asymptotically.
Therefore, the SLP algorithm guarantees jamming of the final packing configurations.

In addition, Our SLP algorithm near the final steps of the process is intimately
related to a linear-programming (LP) protocol to test for
jamming for hard-sphere packings that we previously developed \cite{Do04}. In the LP jamming-test
protocol, the interparticle gaps are maximized subject to the
nonoverlapping conditions. If large gaps can be opened, the packing
under consideration is not jammed; otherwise it is jammed. For our SLP
algorithm, the packing density is driven to a local or global maximum,
which is not only only collectively jammed (since as mentioned earlier,
the descent to maxima necessarily involve collective configurational
motions) but strictly jammed, since the fundamental box
is deforming and shrinking (on average). In general,
if a packing is not jammed, there exist collective particle motions
and boundary deformations that can lead to a higher density.
All of the sphere packings in $\mathbb{R}^2$ and $\mathbb{R}^3$ produced by our SLP algorithm are
tested using the LP jamming-test protocol \cite{Do04}. The packings with a deformable/shrinking
fundamental cell were found to be strictly jammed and those with
an isotropically shrinking fundamental cell were found to be collectively jammed,
which verifies the robustness of our algorithm in producing jammed packings.

\section{Energy Landscape, Inherent Structures (Local Minima) and Global Minima}

\subsection{Energy Landscape Picture}

From a statistical-mechanical  point of view, the solution of the ASC formulation
using a SLP  amounts to searching nearby local minima of the \textit{energy landscape}.
The energy landscape, defined by the energy (objective function), equal to
the negative of the trace of the strain tensor associated with the fundamental cell $F$
of the packing or, equivalently, the negative of the density $\phi$,
is a surface embedded in a $(dN+1)$-dimensional space, where $d$ is the number of
degrees of freedom of a single particle and $N$ is the number of particles in $F$.
A given initial configuration has an energy that corresponds to a point on the landscape.
Starting from this point in this multi-dimensional space,
our SLP algorithm finds a nearby point, which can be reached from the
initial point by a single linear multi-dimensional displacement with a lower energy
(higher density) subjected
to the limits on the size and directions of the displacement as well as the
impenetrability conditions. This procedure is repeated
until a jammed state with a high density is reached,
which can either be a local energy minimum (local density maximum)
or a global energy minimum (global density maximum) depending
on the bound widths on the design variables, as well as the
influence sphere radii $\gamma_{mn}$ for near-neighbor determination.

\begin{figure}[htp]
\begin{center}
$\begin{array}{c@{\hspace{1.5cm}}c}
\includegraphics[height=5.5cm, keepaspectratio,clip]{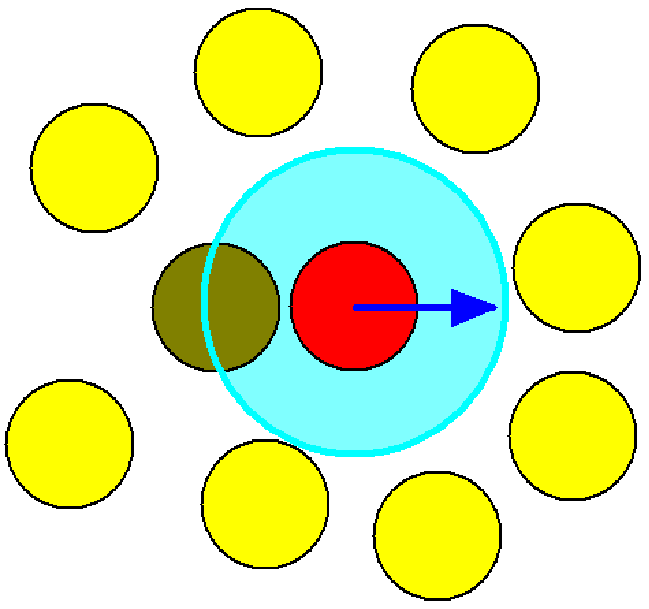} &
\includegraphics[height=5.5cm, keepaspectratio,clip]{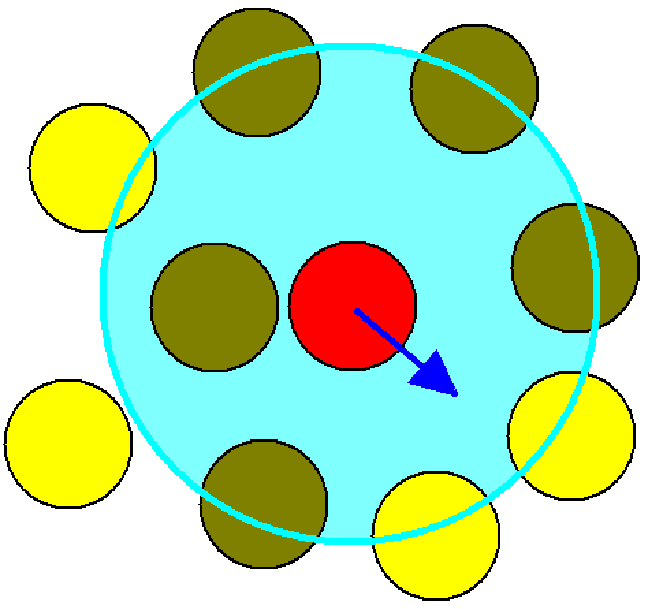} \\
\mbox{(a)} & \mbox{(b)} \\
\end{array}$
\end{center}
\caption{(Color online).  Influence spheres (i.e., effective
interaction range for the central particle in so far as the
nonoverlapping condition is concerned) and the directions of
motion of the particle. (a) Only the nearest neighbor of the
central (red) particle is included in the influence sphere. The central
sphere will move away from its neighbor along a line connecting
their centers, which allows a maximal shrinkage of the
fundamental cell. Such a shrinkage would lead to a local maximum
in the packing density. (b) More near neighbors are included in
the influence sphere. The central particle  moves along a
direction that maximizes its distance from \textit{all} of the
neighbors within the influence sphere. In other words, the
movement of the central particle is coupled with all of the near
neighbors within the influence sphere and allows a maximal
shrinkage of the fundamental cell, which after
a sufficient number of iterations leads to a global maximum
of the packing density.} \label{fig_inflsphere}
\end{figure}

As indicated in Sec.III.A, for monodisperse sphere packings, the radius $\gamma_{mn}$
of the influence sphere for pair $(m.n)$  is identical for each pair of particles.
We call this the \textit{influence sphere} for each
particle (see Fig. \ref{fig_inflsphere}) because only the particles
whose centers are within the influence sphere will affect the
central particle as far as the nonoverlapping conditions
are concerned. For polydisperse packings, $\gamma_{mn}$ is generally
different for each particle pair. However, loosely speaking, one
can still imagine an effective influence sphere around each
particle, which determines the neighbors that affect the central
particle. If the influence sphere (i.e., $\gamma_{mn}$) is
sufficiently small such that only pairs with minimal separation
distance are considered, particle motions are controlled by the local environment (see
Fig.~\ref{fig_inflsphere}(a)), i.e., they move in opposite directions along
the line connecting their centroids. If only particle pairs (or
triplets, etc.) with the minimal separation distance are considered
at each stage of the SLP procedure, the system rigorously follows
the steepest descent trajectory and evolves to the associated
inherent structure. As the influence sphere ($\gamma_{mn}$) becomes
larger, successively larger numbers of  spheres must respect the nonoverlap condition, all
of which affect the motion of the central particle (see Fig.~\ref{fig_inflsphere}(b)). 

Consequently, the influence sphere radius $\gamma_{mn}$ can
be considered to be an effective interaction range. For each
near-neighbor pair, a nonoverlapping condition for that pair is
included in the SLP formulation, which in general will lead to a
different jammed-packing solution  from one in which that
pair of particles are  not considered neighbors of one another. Therefore, the
neighbor pairs effectively interact with each other through the
nonoverlapping condition. A larger $\gamma_{mn}$
corresponds to a longer interaction range, which
enables more directions in the energy landscape to be explored
beyond the one that leads to a local density increase (i.e., a
steepest descent in the energy landscape). We will see in the ensuing
discussion that this feature of the SLP procedure enables it to
find both local and global energy minima (density maxima).
We note that the number of neighbor pairs $n_{pair}$ is a monotonically 
increasing function of $\gamma_{mn}$, namely, $n_{pair} \sim \gamma^d_{mn}$ for any jammed packings in $\mathbb{R}^d$. 
Although a large value of $\gamma_{mn}$ is necessary to produce maximally-dense 
packings, one generally only needs an associated small number of particles in the 
fundamental cell. Therefore, the total computational cost when a large value of $\gamma_{mn}$ 
is employed to generate maximally-dense packings is usually much smaller than that 
associated with a small value of $\gamma_{mn}$ to produce disordered 
packings with a large number of particles per fundamental cell.
Empirically, we find that when $\gamma_{mn} > 4 \overline{D}_{mn}$, the 
SLP algorithm generally leads to the maximally-dense packings, where 
$\overline{D}_{mn} = (D_m+D_n)/2$ is the average diameter of two spheres.

Moreover, one can view this SLP compression process physically as a compression
of a hard-sphere system in a {\it superviscous liquid} subjected to periodic
boundary conditions in the absence of a gravitational field.
Although the immersion of the hard spheres in the superviscous
liquid precludes any collisions between the particles, there is a dynamical interaction
with the system boundary. Thus, the only ``dynamical" parameters of this physical compression
are the deformation parameters of the fundamental cell. For this SLP ``compression" process,
the system moves  efficiently by collective particle motions to nearby denser
configurations.

\subsection{Inherent Structures - Mechanically Stable Local Minima}
\label{inher}

\begin{figure}[htp]
\begin{center}
$\begin{array}{c}
\includegraphics[height=7.5cm, keepaspectratio,clip]{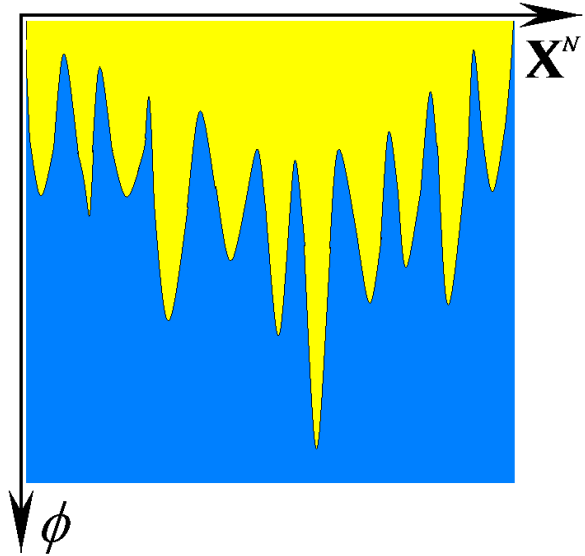} \\
\end{array}$
\end{center}
\caption{(Color online).  A schematic diagram illustrating the
idea of inherent structures for a system composed of $N$ hard
spheres as adapted from Ref. \cite{St64}, though the term
``inherent structure'' was not explicitly used there. The
horizontal axis labeled ${\bf X}^N$ stands for the entire set
of centroid positions, and the density $\phi$ increases downward.
The jagged curve is the boundary between accessible (upper)
configurations (shown in yellow or light gray in the print
version)and inaccessible (lower) configurations (shown in blue or
dark gray in the print version). The deepest point of the
accessible configurations corresponds to the maximal density
packings of hard spheres.} \label{inherent}
\end{figure}

When the influence sphere radius $\gamma_{mn}$ is sufficiently small (e.g.,
$\gamma_{mn}\sim 1.5 \overline{D}_{mn}$) and the bound widths are taken to be sufficiently large
(e.g., $|\epsilon_{ij}|\sim 0.1$ and $|\Delta {\bf x}^{\lambda}|\sim 0.5\overline{D}$,
where $\overline{D}$ is the averaged diameter of all the spheres in the packing),
the SLP solution of the ASC scheme from some initial configurations leads to a mechanically stable
local energy minimum (local density maximum),
which in principle is the {\it inherent structure} associated
with the starting initial many-particle configuration \cite{St82, St85}; see Fig.~\ref{inherent}.
All initial configurations that compress to the
same jammed structure, excluding distinctions between states
that differ by the interchange of identical particles, belong to the same
inherent structure. In two and three dimensions,
the number of distinct inherent structures
scales with the number of particles $N$ in the
fundamental cell, i.e., $\exp(\alpha N)$, where
$\alpha$ is dimensional-dependent constant
of order unity \cite{St82}.

Stillinger, DiMarzio, and Kornegay were the first to introduce the
idea of inherent structures for two-dimensional hard circular
disks \cite{St64}. However, at that time they did not explicitly
use the terminology ``inherent structure.'' Nonetheless, they
indeed proposed a conceptual procedure to produce inherent
structures for hard-sphere systems and applied it to hard disks in
two dimensions. They considered a dilute configuration of $N$ spheres
centered at ${\bf x}_1, \ldots, {\bf x}_N$, and subsequently expanded
the spheres. During such an expansion, there will occur a first
contact between the pair of spheres with the minimal original pair
separation distance. As the spheres are further expanded, one should move
this pair apart (each member at an equal rate along the line joining the centers) 
just to maintain the contact. Other pairs in the system
similarly will come into contact, and will be rearranged by the
same prescription. As the process proceeds, triplets, and larger
sets of spheres will touch, and subsequently should be moved by a
generalization of the pair procedure. These authors
proposed that this procedure can be mapped into  a steepest-descent problem.
Namely, if one has a set of $n$ spheres in contact at ${\bf x}_1,
\ldots ,{\bf x}_n$, the displacements $d{\bf x}_1, \ldots, d{\bf
x}_n$, which maintain contacts under the diameter increase $\Delta
D$, should be selected to minimize the positive definite form
\begin{equation}
\sum_{i=1}^n \left ({\frac{\Delta{\bf x}_i}{\Delta D}}\right )^2
\end{equation}
subject to those constraints.  In particular, these authors
commented that ``for very large $N$, the chance of selecting (at
random) a set of initial positions that would lead to jamming in
the regular close-packed array becomes very small.'' In modern
language, this statement means  that random initial configurations
for the case of identical spheres would lead to  maximally random jammed states for $d \ge 3$, which
are isostatic configurations \cite{fn_isostatic}, with very high probability. However, for
two-dimensional monodisperse circular disk packings, it is well known that random
initial configurations lead to highly ordered crystalline packings
with very high probability \cite{To10}. If an entropic measure
of disorder was employed, such two-dimensional crystalline packings
would be designated to be the most disordered packings or MRJ states,
which is clearly incorrect. This
is one of the many reasons that an
entropic (i.e., occurrence frequency of the same configurations) measure of disorder can be
misleading \cite{To10}.

It is not difficult to see that our SLP algorithm when provided
with the appropriate parameters closely follows the aforementioned
conceptual steepest-descent procedure for obtaining inherent structures. Specifically,
if the non-overlapping conditions are only specified
for nearest-neighbor pairs, triplets etc., i.e., a small
influence sphere radius $\gamma_{mn}$ is used (see Fig.~\ref{fig_inflsphere}),
the SLP algorithm will only move apart those 
spheres that are explicitly considered in the non-overlapping conditions
such that the largest possible compression of the system (consistent with the
non-overlapping conditions and the bounds) can be performed
(through applying the solved strain tensor). In other words,
starting from an initial configuration, the SLP algorithm can lead
to a maximal density increase subject to the constraints
imposed by closest neighbors, following the trajectory of steepest
descent in the energy landscape. This is
an equivalent re-formulation of the Stillinger-DiMarzio-Kornegay
steepest-descent procedure. Importantly, the SLP algorithm is not
limited to steepest-descent mappings. We will see in the next section
that by using relatively large influence spheres,  one can access
unusual inherent structures from random initial configurations.

We note that Zinchenko \cite{Zi94} proposed and implemented
an numerical algorithm conceptually equivalent to the
Stillinger-DiMarzio-Kornegay steepest descent procedure to produce
disordered jammed spheres packings in three dimensions. However,
the Zinchenko algorithm involves the numerical solution of many
different differential equations ($dN$ of them) as well
as the solutions to $dN$ algebraic equations that prescribe
the densification process that allows the particles to swell
while retaining sphere contacts to the extent possible. This makes the algorithm  both
computationally expensive and incapable of generating ordered
packings. Furthermore, O'Hern \textit{et al}. employed a
conjugate-gradient (CG) method to produce jammed packings of soft
spheres interacting with short-ranged repulsive power-law pair
potentials \cite{ohern03}. Starting from a dilute initial
configuration, the spheres are allowed to grow, which cause an
increase of the total energy of the system. Then CG method is used
to relax the system to zero energy. The procedure is repeated
until the total energy cannot be relaxed to zero, and the system
is considered jammed. Although not explicitly indicated, this
algorithm in general produces inherent structures of such 
soft-sphere systems.

\subsection{Energy Minima  - Density Maxima}

Though our SLP algorithm can produce inherent structures, it is
crucial to emphasize that our algorithm is much more general than
the aforementioned conceptual steepest-descent procedure
\cite{St64}, the Zinchenko protocol \cite{Zi94} and the
conjugate-gradient method \cite{ohern03}. 
In particular, including more neighbors associated with any given sphere in the
set of non-overlapping conditions, i.e., increasing the radius of
the influence sphere $\gamma_{mn}$, effectively introduces
``long-range interactions'' (i.e., $\gamma_{mn}\sim 3.5
\overline{D}_{mn}$). Together with smaller bound widths (e.g.,
$|\epsilon_{ij}|\sim 0.01$ and $|\Delta {\bf x}^{\lambda}|\sim
0.05\overline{D}$ (where $\overline{D}$ is the effective diameter
of all of the spheres in the packing), the SLP solution of the ASC
scheme from random initial configuration can lead to unusual
inherent structures, such as a mechanically
stable global energy minimum (global density maximum).


Recall that when the radius of influence sphere ($\gamma_{mn}$) becomes
larger, successively larger numbers of particles must respect the nonoverlap condition, all
of which  affect the motion of the central particle. A
sufficiently large $\gamma_{mn}$ implies that the particles
experience effectively long-range ``interactions'' via the influence spheres, which
enable more directions in the energy landscape to be explored
besides the steepest-descent direction. In other words,
multiple minima can now be explored; and instead of moving to a
local energy minimum (local density maximum) by a steepest descent
trajectory, the system is able to evolve toward the deepest
minimum available until it is in the basin of the global minimum.


\subsection{Unusual Inherent Structures - Low-Density Jammed Packings}

From its definition, it is clear that a inherent structure
is highly dependent on its associated initial configurations. Therefore, an
innovative procedure to generate unusual initial configurations in
principle enables one to obtain unusual inherent structures. Moreover,
there is no reason to limit oneself to random initial
configurations alone. For example, in Ref.~\cite{pnas} diluted MRJ
packings are employed as initial configurations to produce
low-density jammed packings in $\mathbb{R}^3$.

\section{Jammed Disordered and Ordered Packings}

In this section, we employ our SLP algorithm to produce
maximally random jammed (MRJ) packings as well as
maximally dense packings of hard spheres in
$d$-dimensional Euclidean space $\mathbb{R}^d$ for $d=2,3,4,5$ and $6$.
In $\mathbb{R}^1$, there is only a single jammed state,
namely, the integer-lattice packing with unit density.
Therefore, we will not consider this trivial case here.
The characteristics of the packings produced by the SLP algorithm
are compared to those obtained using the LS algorithm, which
verifies the robustness and low-computational cost of our SLP algorithm. Moreover, we
show that the SLP algorithm is superior to the LS algorithm
in several aspects, especially in producing maximal dense packings.

\subsection{Maximally Random Jammed Packings}

\begin{figure}[htp]
\begin{center}
$\begin{array}{c@{\hspace{1.5cm}}c}
\includegraphics[height=6.0cm, keepaspectratio,clip]{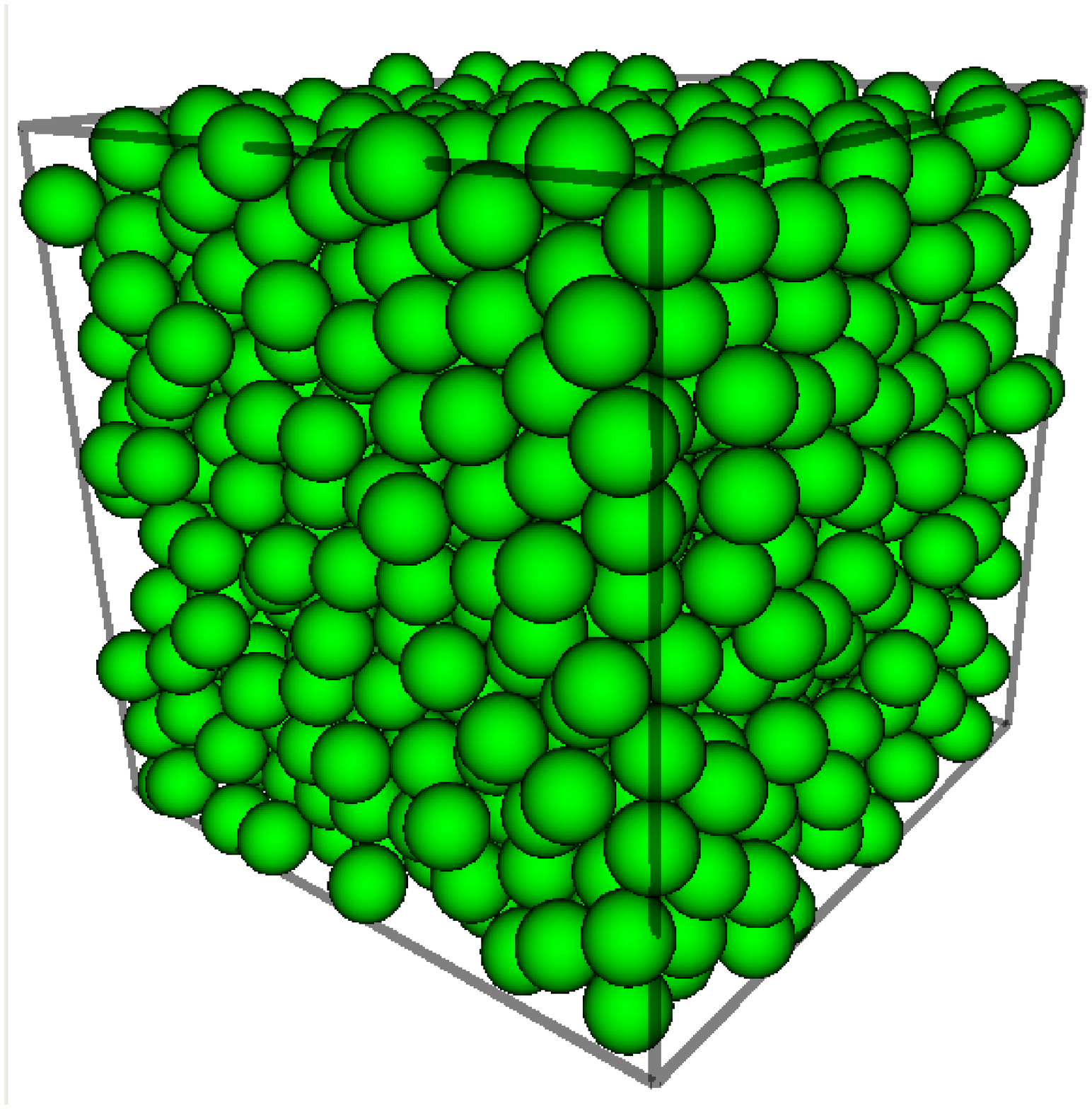} &
\includegraphics[height=5.5cm, keepaspectratio,clip]{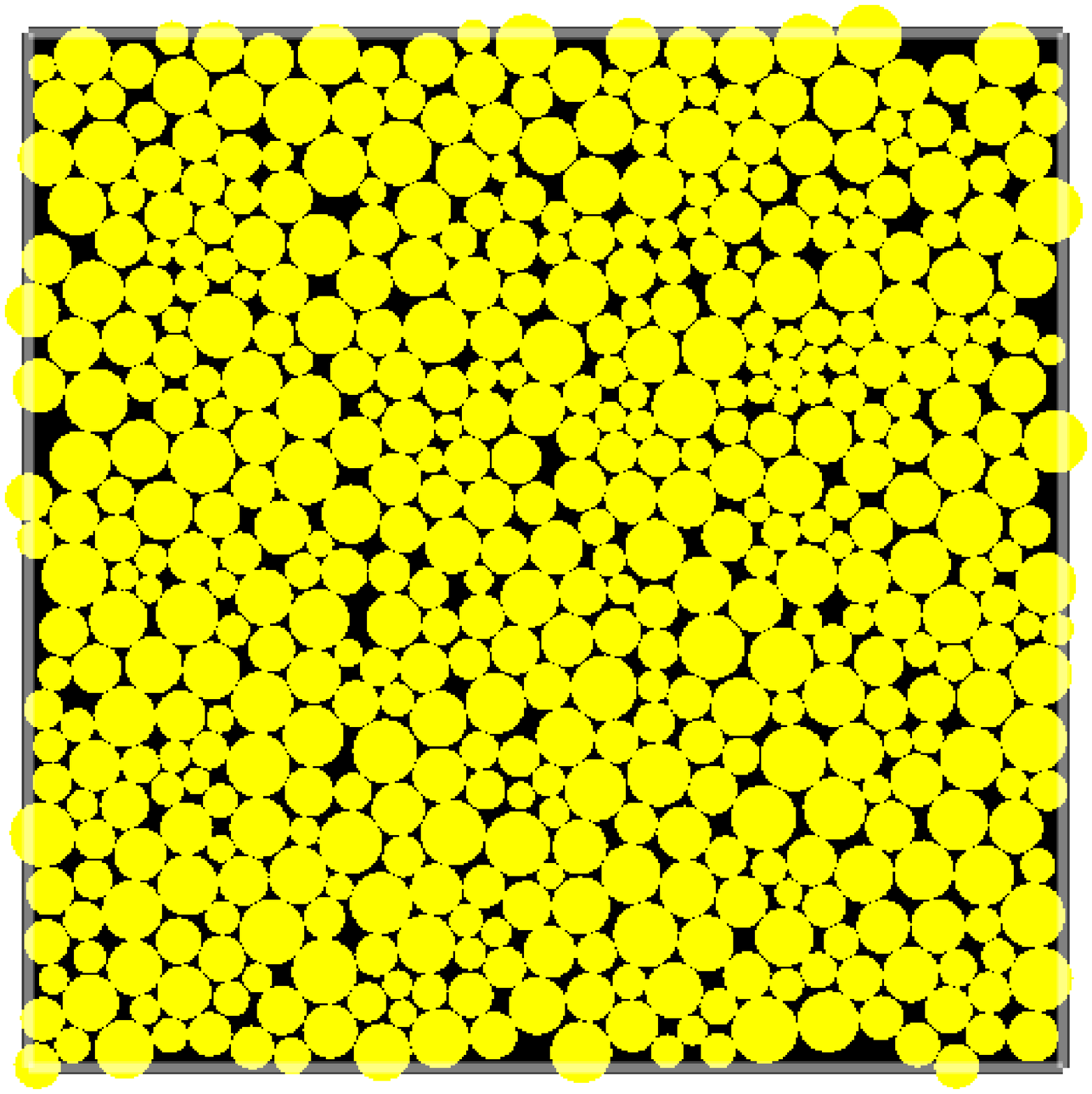} \\
\mbox{(a)} & \mbox{(b)} \\\\
\end{array}$
\end{center}
\caption{(Color online). Maximally random jammed packings in low
dimensions via SLP algorithm. (a) A MRJ packing of 1000 monodisperse
spheres in three dimensions. (b) A MRJ packing of 500 polydisperse
circular disks in two dimensions. }
\label{fig_MRJpacking}
\end{figure}

\begin{figure}[htp]
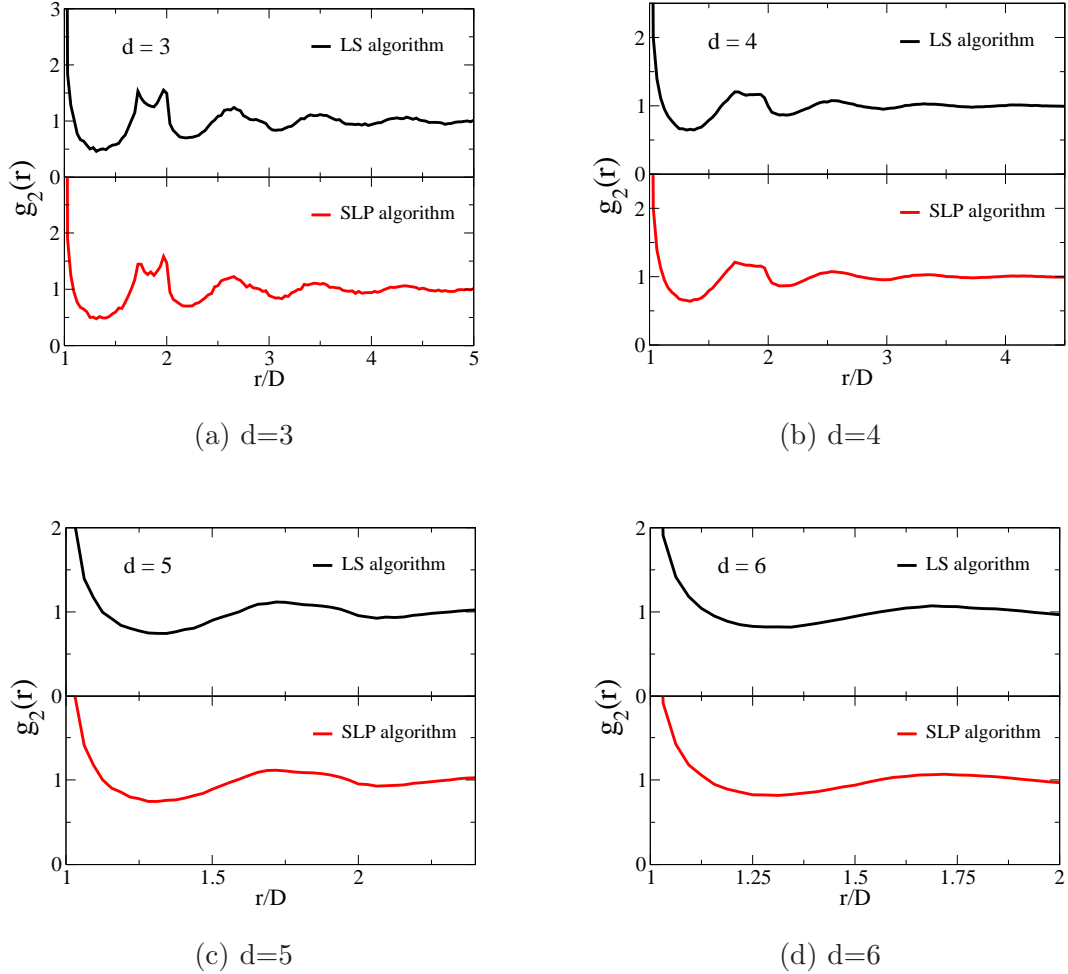

\begin{center}
$\begin{array}{c@{\hspace{1.5cm}}c}
\includegraphics[height=5.15cm, keepaspectratio,clip]{3d_g2.eps} &
\includegraphics[height=5.15cm, keepaspectratio,clip]{4d_g2.eps} \\
\mbox{(a) d=3} & \mbox{(b) d=4} \\\\
\includegraphics[height=5.15cm, keepaspectratio,clip]{5d_g2.eps} &
\includegraphics[height=5.15cm, keepaspectratio,clip]{6d_g2.eps} \\
\mbox{(c) d=5} & \mbox{(d) d=6} \\
\end{array}$
\end{center}
\caption{(Color online). Pair correlation function $g_2$ of
maximally random jammed packings of monodisperse hyperspheres in 
selected dimensions via the SLP and LS algorithm.}
\label{fig_g2}
\end{figure}


Random dilute packings with a density $\phi = 0.05$ are used as
initial configurations to produce MRJ packings for $d=2, 3, 4, 5$ and $6$
using the SLP algorithm.
In two dimensions, polydisperse disks
with diameters uniformly distributed within $[D,~2D]$ are used
to generate MRJ packings because  monodisperse circular disks
have an inevitable propensity to form highly ordered packings in the jamming
limit, as discussed in Sec.~\ref{inher}. In higher
dimensions, monodisperse spheres are used. As indicated in the
previous section, to produce MRJ packings
(which are the maximally disordered inherent structures), a small
influence sphere radius $\gamma_{mn}$ is used such that evolution
of the packing in the energy landscape follows a path of steepest
descent (e.g., $\gamma_{mn}\sim 1.5 \overline{D}_{mn}$,
$|\epsilon_{ij}|\sim 0.1$, and $|\Delta {\bf x}^{\lambda}|\sim
0.5\overline{D}$). The number of particles in the fundamental cell are $N=500, 2000,
4000, 8000, 12000$ respectively for $d=2, 3, 4, 5, 6$. The packing
is considered jammed and the simulation is terminated if the
increase of the packing density is less than $10^{-8}$. Figure
\ref{fig_MRJpacking} shows typical packing configurations in two
and three dimensions.


\begin{table}
\centering \caption{Characteristics of MRJ packings produced using
the SLP and LS algorithms. $\phi$ is the density of final jammed packing
and $f_r$ is the fraction of the rattlers. $t_s$ is the total simulation
time. For LS algorithm, $t_s$ includes the time of simulations with both
the initial large expansion rate and the fine-tuning expansion rate.}
\begin{tabular}{c@{\hspace{0.45cm}}c@{\hspace{0.45cm}}c}
\hline\hline
 &  LS algorithm & SLP algorithm  \\
\hline
d=3    &  $\phi = 0.642\pm0.005$\quad$f_r = 0.030\pm0.003$   &  $\phi = 0.640\pm0.004$\quad$f_r = 0.028\pm0.003$  \\
       &  $t_s = 1.5$ hrs   &  $t_s = 10$ mins  \\
d=4    &  $\phi = 0.460\pm0.006$\quad$f_r = 0.012\pm0.002$   &  $\phi = 0.465\pm0.005$\quad$f_r = 0.014\pm0.002$  \\
       &  $t_s = 4.8$ hrs   &  $t_s = 46$ mins  \\
d=5    &  $\phi = 0.315\pm0.005$\quad$f_r = 0.008\pm0.002$   &  $\phi = 0.310\pm0.005$\quad$f_r = 0.006\pm0.002$  \\
       &  $t_s = 14$ hrs   &  $t_s = 3.2$ hrs  \\
d=6    &  $\phi = 0.201\pm0.005$\quad$f_r = 0.005\pm0.001$   &  $\phi = 0.199\pm0.005$\quad$f_r = 0.005\pm0.001$  \\
       &  $t_s = 193.5$ hrs   &  $t_s = 8.3$ hrs  \\
\hline\hline
\end{tabular}
\label{tab1}
\end{table}

The characteristics of the monodisperse packings are compared to
those obtained using the LS algorithm \cite{To06}, which has been
verified to be a robust protocol to produce MRJ packings. A
two-step procedure is used for the LS algorithm to generate jammed
disordered packings: a large initial expansion rate of $\sim 0.01$
and a small fine-tuning expansion rate of $10^{-(3+d)}$ (where $d$
is the spatial dimension).   A comparison of the packing
density and fraction of rattlers (i.e., movable particles caged by their jamming neighbors)
is given in Table I. Each density $\phi$
and the fraction of rattlers $f_r$ are obtained by averaging over 5 configurations. We note
that the MRJ densities obtained via the LS and SLP algorithms are very
close to each other. We expect that in the infinite-system limit,
the MRJ densities generated via the two algorithms
should be essentially identical to one another. Figure \ref{fig_g2} shows the pair correlation
functions $g_2$ of the packings obtained using both the SLP and LS
algorithm \cite{footnote_3d_g2}. It can be seen clearly that the MRJ packings
produced by our SLP algorithm are consistent with those produced
by the LS protocol across dimensions. Specifically, in
high dimensions, the split-second peak in $g_2$, present for $d=3$,
gets dramatically diminished and oscillations in $g_2$ get
significantly dampened. These findings are consistent with a
recently proposed ``decorrelation principle'' \cite{To06b}
that states that unconstrained spatial correlations in disordered packings
should vanish asymptotically in the high-dimensional limit. The two-dimensional
polydisperse packing possesses a density $\phi = 0.846$ and
approximately $3\%$ of rattlers.

We note that in producing MRJ packings using the LS algorithm, a large expansion
rate is necessary in the beginning of the simulation to drive the system out of
equilibrium but it is undesirable toward the end, which may lead to unjammed
configurations. Therefore, a variable expansion rate is needed and
must be very small at the end of the simulation
to ensure jamming, as we have done here. In our SLP algorithm, the
control parameters are single-valued. Moreover, the SLP algorithm
naturally leads to jammed packings, as we discussed in Sec.III.B, which makes it
superior to the LS algorithm in producing jammed packings.

\subsection{Maximal Density Packings}

In order to further test the SLP algorithm, we apply to see
whether it can recover the densest packings of congruent spheres in two and three
dimensions and densest known congruent sphere packings in $\mathbb{R}^4$,
$\mathbb{R}^5$ and $\mathbb{R}^6$. Recall that we do not consider the
case in $\mathbb{R}^1$, because that corresponds to the trivial solution
of the integer-lattice packing. In $\mathbb{R}^2$ and $\mathbb{R}^3$,
the densest packings of congruent spheres have been proved to be the
triangular lattice \cite{Fe64} and fcc lattice \cite{Ha05}, respectively.
One of the generalizations of the three-dimensional fcc lattice to higher
dimensions is the $d$-dimensional checkerboard lattice $D_d$, defined by taking
a hypercubic lattice and placing spheres on every site at which the
sum of the lattice indices is even, i.e., every other site \cite{Co93}.
The four- and five-dimensional checkerboard lattices
$D_4$ and $D_5$ with densities $\phi=\pi/16=0.6168\ldots$ and
$\phi=\sqrt{2}\pi^2/30=0.4652\ldots$, respectively, are believed to be the densest sphere
packings in those dimensions. For $d=6$, the densest sphere packing is conjectured
to be the ``root'' lattice $E_6$ with density $\phi=\sqrt{3}\pi^3/144=0.3729
\ldots$ \cite{Co93}.

It is important to note that in order to generate perfect crystalline jammed packings
using a  finite simulation cell with periodic boundary conditions, the number of particles
in the fundamental cell must be consistent with its shape.
For example, if a fixed cubic   fundamental cell or rhombohedral
fundamental cell  (associated with the fcc lattice) in three dimensions is used, this
``magic" number of spheres should be $N=4m^3$ and $N=m^3$ ($m=1, 2, 3\ldots$), respectively.
In most cases,
the fundamental cell associated with the densest packing is not known {\it a priori}
and hence one needs to explore a variety of different particle numbers (by carrying out many simulations)
to obtain the maximal density packing with typical protocols that
use a fixed cell. Therefore, it is highly desirable
to explore systematically the shape of the appropriate fundamental cell by employing
a single simulation with variable particle numbers, which our SLP algorithm enables
one to do. In particular, because our SLP algorithm does not simulate the real
many-particle dynamics, a very small value of $N$ can be used, which can significantly reduce
the simulation time. For example, one can choose $N$ to be as small as one,
which reduces the SLP algorithm to an lattice-optimization algorithm.
For the LS protocol, using a very small value of $N$
will introduce systematic errors due to lack of collisions between the particles
within the fundamental cell or because collisions are not properly treated.

\begin{table}
\centering \caption{Characteristics of densest jammed packings produced using
the SLP and LS algorithms. Here $\phi$ is the density of final jammed packing,
$N$ is the number of particles in the fundamental cell, and $t_s$ is the total simulation
time.}
\begin{tabular}{c@{\hspace{0.45cm}}c@{\hspace{0.45cm}}c}
\hline\hline
 &  LS algorithm & SLP algorithm  \\
\hline
d=2    &  $N = 9$\quad$\phi = 0.9068\ldots$   &  $N = 9$\quad$\phi = 0.9068\ldots$  \\
       &  $t_s = 0.1$ mins   &  $t_s = 0.1$ min  \\
d=3    &  $N = 27$\quad$\phi = 0.7408\ldots$   &  $N = 27$\quad$\phi = 0.7408\ldots$  \\
       &  $t_s = 6.5$ mins   &  $t_s = 1.5$ min  \\
d=4    &  $N = 81$\quad$\phi = 0.5608\ldots$   &  $N = 81$\quad$\phi = 0.6168\ldots$  \\
       &  $t_s = 1.6$ hrs   &  $t_s = 4.5$ mins  \\
d=5    &  $N = 243$\quad$\phi = 0.4154\ldots$   &  $N = 243$\quad$\phi = 0.4652\ldots$  \\
       &  $t_s = 47.5$ hrs   &  $t_s = 11$ mins  \\
d=6    &  $N = 729$\quad$\phi = 0.3287\ldots$   &  $N = 729$\quad$\phi = 0.3729\ldots$  \\
       &  $t_s = 283.5$ hrs   &  $t_s = 27$ mins  \\
\hline\hline
\end{tabular}
\label{tab2}
\end{table}

To produce maximal-density packings in $\mathbb{R}^d$, initial
packing configurations with $N=3^d$ spheres in (hyper)cubic
fundamental cells possessing a density lower than the
``equilibrium"  hard-sphere freezing density (Ref.~\cite{Sk06})
for each  dimension (i.e., $\phi = 0.45, 0.4, 0.3, 0.15, 0.06$ for
$d=2,3,4,5,6$, respectively) are used for both the LS and SLP
algorithm. We note that for the current implementation of the LS
algorithm, a deformable fundamental cell program is available for
$d=2$ and 3 \cite{Do05a}; for $d\ge4$, only a fixed cubic
fundamental cell is available \cite{Sk06}. The sphere expansion
rate used to generate a maximally dense  packing in $\mathbb{R}^d$
is $10^{-(3+d)}$. For the SLP algorithm, a deformable fundamental
cell is used in all dimensions. The influence sphere radius is
chosen to be $\gamma_{mn}\sim 3.5 D$), and $|\epsilon_{ij}|\sim
0.01$ and $|\Delta {\bf x}^{\lambda}|\sim 0.05D$ are used. The
final densities produced by  the two algorithms and the
corresponding computational times to produce the jammed states are
compared in Table II. It can be seen that the SLP algorithm
successfully reproduces the densest known packings in all
dimensions; while our implementation of the LS algorithm only was
able to reproduce the maximal-density packings in $\mathbb{R}^2$
and $\mathbb{R}^3$. Although the LS algorithm does not generate
perfect crystals for $d=4,5$ and 6, the high densities, it
produces  suggest that these packings contain very large
crystallized regions. We have also used $N=128$ in $\mathbb{R}^4$
and $N=512$ in $\mathbb{R}^5$ for the LS algorithm, which are the
magic numbers for cubic fundamental cells in four and five
dimensions, respectively, as well as an expansion rate of
$10^{-10}$. In $\mathbb{R}^4$, the packing converges to the
$D_4$-lattice packing with $\phi = 0.6168\ldots$; while in
$\mathbb{R}^5$, the final packing is partially crystallized with
$\phi = 0.4392\ldots$.

These results clearly demonstrate the advantage of our SLP algorithm in producing
maximally dense sphere packings, i.e., the SLP algorithm is both much more computationally efficient and
more robust in finding the densest known sphere  packings than the LS algorithm. In particular, the adaptive fundamental cell is crucial for obtaining
perfect maximal density packing structures.
In addition, we have tried a variety of particle numbers in the fundamental cell
(e.g., $1\le N\le 2^d$ for $\mathbb{R}^d$) in four, five and six dimensions and have not obtained any
packings possessing a larger density than the densest known packings. This provides
numerical support that the densest known packing in these dimensions may indeed be the
conjectured maximal-density packings.

We note that very recently Kallus, Elser and Gravel \cite{Ka10}  applied  a periodic
``divide'' and ``concur'' (PDC) algorithm to generate maximally dense packings
of hard particles. In this variant of the  PDC algorithm, they follow
Ref. ~\cite{To09} by allowing the fundamental cell to deform
and change shape during the simulation. Interparticle
impenetrability constraints are relaxed in the beginning of the simulation
such that overlapping intermediate configurations are allowed.  Towards the end
of the simulation, all overlaps are systematically
removed. Among other applications, these authors
applied the PDC algorithm to successfully reproduce the densest known
lattice sphere packings (not necessarily the densest packings)
and the best known lattice kissing
arrangements in up to 14 and 11 dimensions, respectively, since they
were limited to  using only a small number of
particles per fundamental cell. Although the PDC algorithm is
efficient in producing dense lattice sphere packings, it does not appear
to be capable of generating \textit{jammed} sphere packings with a
diversity of disorder and density that can be produced via either the SLP
or LS algorithm.


\section{The Order Map}
\label{order}


\begin{figure}[htp]
\begin{center}
$\begin{array}{c}
\includegraphics[height=7.5cm, keepaspectratio,clip]{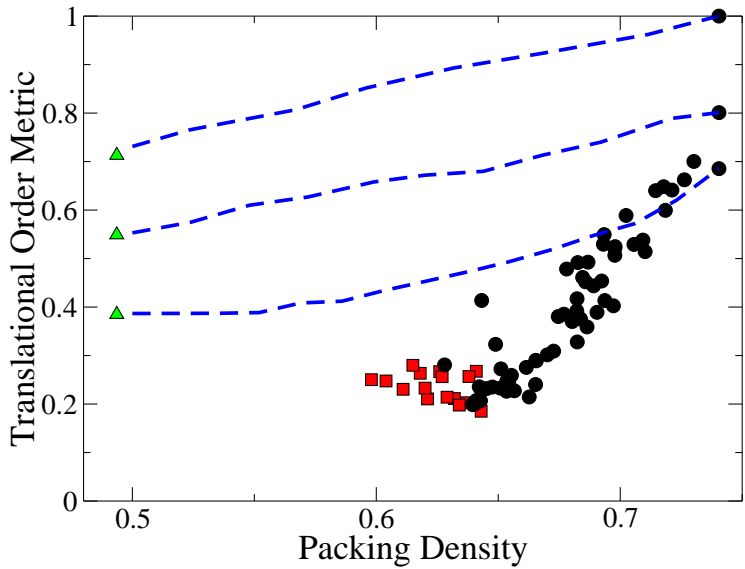} \\
\end{array}$
\end{center}
\caption{Order map for strictly jammed packings of congruent spheres in
three dimensions with $N\sim1000$ generated using the SLP
algorithm: translational order metric $\mathfrak{T}$ versus
packing density $\phi$. The black circles are data produced
via the SLP algorithm using random initial conditions. While the density
in this data set ranges essentially continuously from $\phi \approx 0.64$, which coincides
with that of the MRJ state, up to the maximal density of $\phi=\pi/\sqrt{18}=0.7408\ldots$,
the degree of order at fixed density is variable. The red squares
are data corresponding to disordered jammed packings
with anomalously low densities, with density as low as
$\phi \approx 0.595$, which were generated via the SLP protocol using special
initial conditions involving diluted MRJ packings, as described
in the text. The dashed blue curves represent the spectrum of packings
generated by randomly filling the vacancies in three of the tunneled
crystal packings down to the density $\phi=\sqrt{2}\pi/9=0.49365\ldots$ (green
triangles), each of which leads to the corresponding maximal-density Barlow
packing [8] (with $\phi=\pi/\sqrt{18}=0.7408\dots$)}.
\label{fig_ordermap}
\end{figure}

Besides the capability of generating  MRJ packings and maximally dense
jammed packings efficiently and robustly, the SLP
algorithm has the capacity to produce jammed sphere packings of
with a wide range of densities and degrees of
disorder/order, i.e., a diverse set of inherent
structures. Here we illustrate this capability
by focusing on identical spheres in three dimensions. By tuning the bound widths and the influence
sphere radius $\gamma_{mn}$, the density at which the packing
is jammed and its degree of disorder can be controlled
to a great degree.  In particular, the ranges of
the parameter values used are as follows: $\gamma_{mn}\in (1.5
{D}_{mn}, 3.5{D}_{mn}$), $|\epsilon_{ij}|\in
(0.01, 0.1)$ and $|\Delta {\bf x}^{\lambda}|\in(0.05{D},
0.5{D})$ (where ${D}$ is the diameter of the
spheres in the packing). In general, a larger value of $\gamma_{mn}$
and smaller bound widths allow for  a larger number of
particles to be effectively coupled to one another, which in turn
leads to a higher jammed packing density. By continuously varying the
values of the control parameters, a spectrum of packing density
$\phi\in[0.64,~0.74048...]$ can be obtained when random dilute
packings are used as initial configurations. This density range
is also achievable by the LS algorithm, but we shall see
that disordered jammed packings with densities significantly lower than $0.64$ can be
generated using the SLP algorithm.


Typical configurations of such packings with $\phi\in[0.64,~0.74048...]$
and $N\approx 1000$ are
mapped onto a density-order-metric diagram (black circles). The
diagram or \textit{order map} emphasizes a ``geometric-structure''
approach to analyze packings by characterizing individual
configurations, regardless of their occurrence probability
\cite{To10}. Here the translational order metric $\mathfrak T$
\cite{To00} is used to quantify the order of the packings (other
order metrics yield equivalent results \cite{To00, Ka02}).
The translational order metric $\mathfrak T$ \cite{To00} is
defined as
\begin{equation}
\label{trans_order}
\mathfrak{T} =
\left|{\sum_i^{N_s}(n_i-n_i^{ideal})/\sum_i^{N_s}(n_i^{FCC}-n_i^{ideal})}\right|,
\end{equation}
where $n_i$ is the average occupation number for the shell $i$
centered at a distance from a reference sphere that equals the
$i$th nearest-neighbor separation for the open FCC lattice at that
density and $N_s$ is the total number of shells for the summation;
$n_i^{ideal}$ and $n_i^{FCC}$ are the corresponding shell
occupation numbers for an ideal gas (spatially uncorrelated
spheres) and the open FCC lattice. For a completely disordered
system (e.g., a Poisson distribution of points) $\mathfrak{T} = 0$, whereas
$\mathfrak{T} = 1$ for the FCC lattice.


Moreover, we have produced jammed disordered packings
with densities as low as $0.595$, which is an anomalously low
value,   using the
SLP algorithm from special initial configurations. Specifically,
a small fraction ($f_s = 0.1\% - 2.5\%$) of particles are removed
from MRJ packings which are obtained using the SLP algorithm to
produce unjammed initial configurations. The remaining unjammed
spheres are then compressed to a jammed state using the SLP
algorithm. Such a procedure is repeated $n_r = 5 - 10$ times
before a lower limit on the final jammed packing density  $\phi$ is reached. Low-density jammed
packing with densities in the range  $\phi\in[0.6, ~0.64]$ have been produced (removing
the rattlers in the packing results in a slightly lower density,
$\phi \approx 0.595$). These packings are mapped onto the
$\phi$-$\mathfrak{T}$ diagram (red squares). We note that a
similar procedure was employed in Ref.~\cite{pnas} to produce
low-density jammed packings. However, there the LS algorithm instead of
the SLP algorithm was used to generate the initial configurations. 
Moreover, we would like to emphasize that the low-density jammed 
packings produced by the SLP algorithm are completely unrelated to 
the so-called ill-defined ``random loose packings'' of hard spheres, 
which have been shown to be not even strictly jammed \cite{Do04}.

``Tunneled" crystal packings (green triangles) \cite{To07} and maximal
density packings (fcc, hexagonal close-packed, and their
stacking variants called the Barlow packings) are also shown on
the order map. The blue dashed curves show the spectrum of packings
generated by randomly filling the vacancies in the tunneled
crystal packings, which leads to the corresponding maximal density
packings. These dashed curves were first reported in Ref.~\cite{pnas}.
The tunneled crystals have density $\phi=\sqrt{2}\pi/9=0.49365\ldots$
are currently the best candidates for the lowest density
strictly jammed packings in three dimensions. bf Each sphere
in any of the tunneled crystals contacts exactly 7 others
and therefore such packings are {\it hyperstatic}.

We note that for comparison to existing literature, the data presented
in Fig.~\ref{fig_ordermap} for the order map include the rattlers, which constitute a small percent of
spheres near the MRJ state (e.g., $2 \sim 3 \%$). The fraction of
rattlers decreases as the packing density increases or decreases from
MRJ state along the lower boundary of jammed states (in either direction in density). In addition,
consistent with the results reported in Ref.~\cite{pnas}, we find that
the average contact number $Z$ per particle increases from the
isostatic value of 6 at the MRJ state as the density increases or
decreases from the MRJ-state value along the lower boundary of jammed
states. In particular, at $\phi \approx 0.6$ the average contact
number $Z \approx 6.3$. It is expected that $Z$ will increase to a
contact value per particle of 7 as $\phi$ is decreased to the tunneled-crystal
value $\phi = 0.49365\ldots$. On the other hand, as $\phi$ is increased from the MRJ value,
the average contact number $Z$ continuously increases from 6 all the way up to
12 for the hyperstatic Barlow packings with a density of
$0.74048\ldots$. Thus, moving off the MRJ state along the
lower boundary of jammed states is associated with
an increase in the average contact number $Z$ and the
degree of order.

\section{Discussion and Conclusions}

In this paper, we have proposed and implemented a sequential-linear-programming
procedure to solve the adaptive-shrinking-cell optimization problem to generate
jammed sphere packings. The SLP procedure is particularly suitable and natural for
sphere packings with a size distribution because
the objective function and impenetrability constraints can be exactly linearized
and the final state is ensured to be jammed in principle.
We have shown that the SLP algorithm can produce robustly a wide spectrum
of jammed sphere packings in $\mathbb{R}^d$ for $d=2,3,4,5$ and $6$
with varying degrees of disordered and densities.
In particular, we applied the algorithm to generate various disordered packings
as well as the maximally dense packings for $d=2,4,5$ and 6.
Moreover, we showed that our SLP algorithm can produce with high probability a variety of
collectively jammed packings with a packing density anywhere in the wide range $[0.6, 0.7408\ldots]$
in three dimensions, which supports the view that there is no universal
jamming point that is distinguishable based on the packing density alone \cite{Ka02,To10,pnas}.
Our jammed sphere packings are characterized
and compared to the corresponding packings generated by the well-known and versatile
Lubachevsky-Stillinger molecular-dynamics packing algorithm.
Compared to the LS procedure, our SLP protocol is able to
ensure that the final packings are truly jammed,
produces disordered jammed packings with anomalously low
densities, and is appreciably more robust and computationally
faster at generating maximally dense packings, especially as the
space dimension increases. Therefore, the SLP algorithm for producing jammed sphere
packings retains the versatility and advantages of the
LS algorithm while improving upon its imperfections.

An important feature of the SLP algorithm is that it can produce
a broad range of inherent structures (locally maximally dense
and mechanically stable packings), besides the usual
disordered ones (such as the maximally random jammed state) with very small computational cost
compared to those of the best known packing algorithms
by tuning the radius of the {\it influence sphere}.
This is to be contrasted with other packing protocols, such as the
conjugate-gradient method for soft three-dimensional spheres \cite{ohern03} and
the Zinchenko algorithm for three-dimensional hard spheres \cite{Zi94}, which can only 
lead with high probability to the MRJ-like inherent structures from random initial
configurations. These protocols are limited in the sense that the trajectory
from one point to another on the energy landscape is strongly limited, i.e.,
only those directions corresponding to local steepest descents can be chosen. This makes
the final packings strongly dependent on the random initial configurations. As we have shown,
exploring other directions in the energy landscape can lead to a variety of jammed packing structures
with variable disorder and density even if the same random initial configuration is used.
In addition, it is not clear that these other packing protocols
would lead to inherent structures with the same level of order/disorder, since
no order metrics have been explicitly measured for the resulting jammed packings.
This emphasizes a major point made in Refs.~\cite{Ka02,To10,pnas}, namely,
although the diversity of packing protocols have yet
to be fully explored, there is already substantial
evidence that available algorithms, including
the one reported in this paper,  can produce jammed packings at the same density
but with a wide range of order/disorder.

A straightforward generalization of the current SLP procedure
for hard-sphere packings in Euclidean space $\mathbb{R}^d$ is to devise a linear-programming
protocol for packing spheres in curved spaces, such as the surface of a
sphere. In such cases, additional constraints on the sphere displacements
are necessary to keep them on the spherical surface,
but these constraints are simple to incorporate.

One advantage that the LS packing protocol has over any SLP
algorithm is in producing jammed packings of smoothly shaped
nonspherical particles (e.g., ellipsoids and superballs) because
the nonoverlap functions are nonlinear \cite{Do05a,Do07,Ji08,Ji09} and linearization
of the constraints is no longer rigorous and hence
jamming cannot be guaranteed. However, in such
instances, the ASC scheme can be solved using Monte Carlo
methods, as was done for hard-polyhedron packings \cite{To09},
or by using nonlinear programming methods \cite{Ru06}.
When the particle shape deviates only slightly from a perfect sphere, 
a LP solution could still be possible. Moreover, if one is interested in the true dynamics
that leads to jamming, the LS algorithm is clearly preferred
over the SLP protocol, which is a deterministic algorithm that
is incapable of capturing the real dynamics. However,
if one is simply interested in generating jammed packings
without regard to history, which the geometric-structure point of view
advocates \cite{To00,To10,pnas}, then the SLP protocol is preferable.

Finally, we note that it is highly desirable to devise packing protocols
that targets the density or other packing characteristics while ensuring
jamming. It is natural to formulate such an inverse problem
as an optimization problem  \cite{To10b} and to solve it for sphere packings using  a LP solution procedure. In future work,
we will generalize the current SLP algorithm to create even lower
density jammed states of hard spheres than have been able
to produced thus far. In three dimensions,
this could enable the large gap without any  jammed states
between densities $\sim 0.49$ and $\sim 0.6$ with relatively
low order metric values (see Fig.~\ref{fig_ordermap}) to be filled
in.


\begin{acknowledgments}
We thank Aleksandar Donev for helpful comments on the manuscript.
This work was supported by the Division of Mathematical Sciences
at the National Science Foundation under Award Number DMS-0804431
and by the MRSEC Program of the
National Science Foundation under Award Number DMR-0820341.
\end{acknowledgments}


\begin{thebibliography}{44}
\bibitem{rate}
The sphere expansion rate $\gamma$ is defined as $\gamma =
\frac{1}{D}\frac{dD}{dt}$, where $D$ is the diameter of the
sphere.
\bibitem{pnas-f}
In Ref.~\cite{pnas}, it was shown that jammed packings with a very narrow density
distribution centered at any $\phi$ over a wide range $\phi \in [0.6,~0.7408\ldots]$
with variable disorder can be generated in thre dimensions. These results 
as well as those reported here support the view that
there is no universal jamming point that is distinguishable based on the
packing density alone and occurence frequency. These results stress the importance of 
the ``geometric-structure'' approach to analyzing packing problems [14].
In Ref.~\cite{pnas}, the SLP algorithm was employed to produce
disordered jammed sphere packings in three dimensions with anomalously low densities. However,
the algorithmic details of the SLP solution were not provided in that paper.
Here we also use the SLP algorithm to generate disordered jammed sphere packings
with anomalously low densities but using only initial conditions produced
by the SLP algorithm. Additional details are provided in Sec.~\ref{order}.
\bibitem{footnote2}
In practice, the relative coordinates (rather than the global coordinates)
are used in the algorithm due to the changing sphere positions
with respect to the adaptive fundamental cell.
Note that the terminology ``global" coordinate was not explicitly
used in Ref.~\cite{To09}, even though the equations
for the global positions of particle centers were
explictly given.
\bibitem{fn_isostatic}
A packing is isostatic if it possesses the minimal number of contacts
for a particular jamming category.
In large-packing limit, this implies that the average contact per
particle in a strictly jammed packing is equal to 2$d$, where $d$ is the spatial
dimension [14, 45].
\bibitem{footnote_3d_g2}
We note that there is a slight difference between the pair correlation functions for the 
three-dimensional MRJ packings produced by the LS and LSP algorithms, 
which appears to be primarily due to the statistical noise. 
The corresponding translational order metrics of the packings are measured to be 
$\mathfrak{T}_{LS} = 0.197$ and $\mathfrak{T}_{SLP} = 0.194$, which indicate that the 
packings virtually have the same degree of disorder.
\end{thebibliography}
\end{document}